# LIMB PREFERENCE IN THE GALLOP OF DOGS AND THE HALF-BOUND OF PIKAS ON FLAT GROUND

**R. HACKERT, L.D. MAES, M. HERBIN, P.A. LIBOUREL, A. ABOURACHID**
UMR 7179, Dpt EGB, Museum National d'Histoire Naturelle, Paris, France
57 rue Cuvier CP 55 75231 Cedex

**Abstract -** During fast locomotion — gallop, half bound — of quadruped mammals, the ground contact of the limbs in each pair do not alternate symmetrically. Animals using such asymmetrical gait thus choose whether the left or the right limb will contact the ground first, and this gives rise to limb preference. Here, we report that dogs (Mammalia, Carnivora) and pikas (Mammalia, Lagomorpha) prefer one forelimb as trailing limb and use it as such almost twice as often as the other. We also show that this choice depends on the individual and is not a characteristic of the species, and that the strength of the preference was not dependent on the animal's running speed.

## INTRODUCTION

The expression of brain lateralization has been observed in behaviours as different as feeding, manipulation, and communication in several families of vertebrates. Limb preference is a subset of this expression and has been documented in birds and in most families of tetrapods including toads, anurans and mammals (for review in Vallortigara et al., 1999). In some species, sex was identified as a determinant of the direction of the preference at population level (in dogs: Wells, 2003; Quaranta et al., 2004; Poyser et al., 2006; in cats: Tan et al., 1990) but not of the strength of preference. The strength of preference is thought to be task dependent and increases as tasks require more cognitive function (Fagot and Vauclair, 1991) and coordination (Hopkins, 1995).

Curiously, limb preference during locomotion has not been extensively explored, although locomotion is the primary function of limbs. Malashichev (2006) observed five species of anuran and found a limb preference in those that exhibit an "alternated limb locomotion or other unilateral limb activity". He suggested that the degree of lateralization in the motor response depends on the mode of locomotion used by a species. In the asymmetrical gaits of mammals– i.e. gallop, bound, and half-bound — the fore and hind pairs of legs make contact with the ground alternately. During gallops or half-bounds, the motions of the two limbs of a pair are not symmetrically alternated, such that one limb of each pair touches the ground first and is called the *trailing* limb, the other one is called *leading* limb. The trailing and leading limb of each pair cannot be systematically assigned to the left or right limb because during locomotion animals switch trailing and leading limbs from time to time. In their study of the gallop of four race horses, Deuel and Lawrence (1987) compared the kinematics of trailing and leading limbs. All the horses preferred the right limb as trailing limb. Moreover, the duration of the stance phase of the trailing forelimb differed significantly according to whether it was the right or the left limb. Walter and Carrier (2007) observed that four of the six dogs in their study preferred one limb as the trailing limb during gallop. Finally, Hook and Rogers (2002) found that marmosets land preferentially with the right limb and that the choice of this limb is correlated with the leading limb used during leaping and walking.

There have been few quantitative studies of the direction or the strength of limb preference during animal locomotion. We therefore investigated this issue and changes in laterality with speed in dogs and pikas, two mammals with different morphologies, behaviours and gaits. Dogs are medium-sized digitigrades that trot over a large range of speeds before switching to galloping at high speeds. In contrast, pikas are small plantigrades that switch to half-bound locomotion for all speeds faster than walking (Gambaryan, 1974; Fischer and Lehmann, 1998).

## ANIMALS AND METHODS

We studied five dogs (*Canis familiaris*, Mammalia: Carnivora) and four pikas (*Ochotona rufescens*, Mammalia: Lagomorpha). The dogs were male Belgian shepherd malinois of similar size (withers height=0.61±0.05 m, BW=28.3±2.0 kg); they belonged to the French army and were thus healthy, obedient, and used to energetic exercise. Pikas (withers height=0.05 m, body length=0.16 m, BW=150-200 g) were reared at the IPBS (CNRS, Toulouse) by A. Puget and were kept at the Motion Lab, University of Iena, Germany; they were reared and used in conformity with German animal welfare regulations.

### *Methods*

Dogs were studied while running outdoors along a 1 m x 12 m long carpet graduated each 10 cm for stride length calculations. They were filmed with a high speed camera (Basler A504K, highland Illinois, USA) positioned 8 m from the left-hand side of the carpet and driven at 200 Hz. The camera was also driven to maintain the dog in the field of view (about 3 m). Under outdoor conditions – in contrast to treadmill —the animal's speed is not controlled and the collected locomotor cycles depend largely on the performance and motivation of the dog. The dog handler ran the distance at five different speeds to allow the dogs to gallop at different speeds. Dogs were all lead on a leash by the handler on the right-hand side. The two meters long leash was light and long enough to avoid interfering with the dog's locomotion. To study speeds faster than human running speed in a second series of records, dogs were also encouraged to run freely towards a ball at the opposite end of the same carpet. We used this second series of records to test whether the handler had an influence on the direction of the preference. The direction of preference for each dog was the same in trials with and without the leash. Trials where the leash was bent or where the animal looked at the handler (very rare because galloping quadruped mammals lock the degree of freedom of the neck) were excluded from our data set. Animal speed was calculated from stride length and stride duration. Speeds were between 2.2 m/s and 10.3 m/s and each sequence contained 4-8 locomotor cycles according to speed. However, each dog did not cover exactly the same range of speeds.

Pikas ran on a treadmill and running speed was imposed on the animals. They were habituated to the manipulation and to the running environment for 15 minutes per day for two weeks. High speed video films were recorded with a CAMSYS® camera driven at 400 Hz and equipped with a Fujinon zoom-lens (1.2/75) adjusted to cover one body length. Sequences of 4s were recorded at speeds between 1.2 m/s and 2.2 m/s, a range of speeds comfortably achieved by the four animals. The speed was increased by 0.05 m/s between consecutive sequences. Each sequence included between 15 and 30 locomotor cycles. Animal were allowed rest for 3 minutes between the running sequences.

### *Processing films and statistics*

We noted the limb used as the trailing limb in each locomotor cycle recorded on each film sequence; films were each scored twice independently and readings compared. There was no disagreement associated with the determination of the trailing limb.

However, assuming that the left/right distribution converges on a particular value as the number of cycles analysed increased, the truncation of data acquisition after any given number of cycles would limit accuracy. The power analysis module of statistical software (STATISTICA v.6, *Statsoft*) was used to determine the accuracy according to the sample size.

We grouped our data into three classes of speed: slow (S), medium (M) and fast (F) gallop (or half-bound). As the range of speed covered depended on the motivation of the dogs to perform the exercise, speed classes were adapted to the performance of each dog by dividing the range of speed into three equal intervals for each animal; thus, the classes differed between individuals. The limits of the classes are given in Fig.1. For pikas, the three classes each had the same width [1.2-1.5] m/s, [1.5-1.8] m/s and [1.8-2.2] m/s; these classes have biological relevance as pikas introduce a first aerial phase at about 1.5-1.6 m/s and then a second at around 1.8-1.9 m/s (Fischer and Lehmann, 1998; Hackert, 2003).

The significance of our binary (right/left) data was tested for each speed class using a $Chi^2$ test (1 degree of freedom) to detect a significant difference ($p<0.05$) from a 50:50 distribution. Classes were then compared two by two using a $Chi^2$ test ($p<0.05$, $H_0$= "distributions are different").

A handedness index (HI) was calculated by subtracting the total number of left-trailing touch-downs (L) from the total number of the right-trailing touch-downs (R) divided by the total number of trailing touch-downs: $(R - L)/(R + L)$. The resulting values, ranging from -1.0 to 1.0, provide a score for each animal's hand preference on a continuum from strongly left-handed to strongly right-handed. The absolute value of the HI (ABS-HI) indicates the strength of hand preference, irrespective of its direction (see Hopkins, 1995).

## RESULTS

Left- and right-handed individuals were found among dogs and pikas. In each speed class the 10 animals significantly preferred ($p<0.05$) one limb, except pika 4 and the slow classes of pika 1 and of dog 3 (fig. 1). The direction of preference remained the same in the three speed classes for each animal. Comparing the strength of limb preference of speed classes two by two, significant ($p<0.05$) differences were observed only between the slow class of pika 1 and the medium and fast classes, and between the fast class of pika 2 and the slow and medium classes. Thus, we observed a constant strength of limb preference in 8 of 10 animals, including both dogs and pikas. The strength of preference varied between individuals and was generally stronger in dogs (ABS-HI from 0.31 to 1, mean ABS-HI= 0.65) than in pikas (ABS-HI from 0 to 0.46, mean ABS-HI= 0.3).

## DISCUSSION

Lateralized behaviour in dogs has been described (Tan, 1987; Wells, 2003; Quaranta at al. 2004; Van Alphen et al., 2005; Poyser et al. 2006; Branson and Rogers, 2006) but not as concerns locomotion. Our study shows that paw preference is present during asymmetrical locomotion in two quadrupeds: dogs and, for the first time, a lagomorphian.

Previous studies report an influence of sex on the direction of the lateralization at the population level in dogs (Quaranta et al., 2004) but not on the strength of the lateralization (Wells, 2003). Our group only included adult males and was equally divided into right- and left-pawed dogs, but the small number of subjects studied prevents conclusions at the population level.

The source of asymmetry in gallop and half-bound locomotion corresponds to a functional differentiation between the trailing and leading limbs of a pair. Indeed, Walter and Carrier (2007) found significant differences in the peak vertical forces of galloping dogs in the hind but not in the fore pair, a larger decelerating impulse in the leading than trailing forelimb and a more accelerating impulse in the leading than trailing hind limb. Hence, the specific exercises of each limb may lead to a specific muscular fatigue. To minimize this muscular fatigue, animals often switch between trailing and leading limbs when galloping and half-bounding. The strength of preference observed thus depends on the number of switches and the number of consecutive cycles recorded in a locomotor sequence. In our study of dogs, sequences included 4-8 consecutive locomotor cycles; for pikas locomotor sequences involved 15-30 consecutive cycles, potentially giving rise to more fatigue and consequently more switches. This methodological difference may have contributed to the lower strength of preference observed in pikas.

Ecological factors certainly influence the strength of preference during gallop in a species. Using the data published by Deuel and Lawrence (1987) concerning a group of four ridden race horses, all left footed (15 cycles recorded per horses), we calculated a mean HI of 0.32 (HI from 0.12 to 0.50). This is slightly higher than the value we obtained for pikas but lower than that for dogs. This indicates that the strength of preference does not correlate strongly with animal size. Possibly, the type of locomotion influences the strength of lateralization: pikas, a model for many small mammals, do not plan their trajectory and therefore discover the relief at the last moment. Manoeuvrability, i.e. the ability to suddenly turn left and right, and reactivity are most important for prey species like pikas. In contrast, the locomotion of dogs and other carnivores is goal directed because they gallop when pursuing prey; their locomotion is more cognitive. These two types of locomotion correspond to two different tasks —escaping or hunting — and define a gradient of cognitive involvement (Fagot and Vauclair, 1991). On this gradient, horses like pikas are prey, but with a much larger field of view and so there is a larger cognitive input. Indeed, comparing species involves defining more precisely the types of locomotion they use, and thus the types of task they perform.

In their study of manual laterality in non human primates, Fagot and Vauclair (1991) developed a framework to classify tasks. They defined a gradient for the strength of lateralization according to the complexity of the task. Complexity is here defined in terms of spatiotemporal precision of the movement and novelty of the situation (inducing cognitive functions). They thus opposed complex *high level* tasks to *low level* tasks including familiar practised activities that should require less cognitive involvement.

Different tasks have been given to dogs to explore the strength of their paw preference: for example retrieving a chocolate from a can (mean ABS-HI=0.39; Wells, 2003), and holding a hollow cylinder filled with food (Kong test; Branson and Rogers, 2006). In the latter experiment the ABS-HI values were between 0 and 0.8, with most falling between 0.1 and 0.5; we estimate that the mean ABS-HI value was around 0.35-0.40, a value in accordance with the 0.39 found by Wells for the other task. The slightly lower value found for removing an adhesive tape from the nose (mean ABS-HI about 0.30; Quaranta et al., 2004) appears natural since the movement is directed towards a body part – the nose - and is thus more stereotyped i.e. familiar comparable in some extent to cleaning movement. In contrast, removing a blanket from head required hand-head coordination, and is thus a more complex task and this HI was found higher (mean ABS-HI=0.5; Wells, 2003). The mean ABS-HI values for all these idiomotive tasks falls between 0.3 and 0.5 despite large variations at the individual level and notable differences in the composition of the groups of animals. For paw lifting on command, Wells found, by contrast, a very high degree of lateralization (mean ABS-HI=0.8) but this may have been due to learning reinforcement effects affecting performance. Thus, asymmetrical locomotion in dogs (mean ABS-HI=0.65) seems to be subject to a strong preference, although there are methodological issues to be considered when interpreting this finding.

Where does asymmetrical locomotion fall on the gradient defined by Fagot and Vauclair? Cyclic locomotion on flat terrain is by nature the most stereotyped and practised movement and consequently belongs, *a priori*, to the class of low level tasks. However, galloping *stably* sets constraints on the animal's motion: touching down at the appropriate instant with the trailing limb improves the dynamic stability of running (Seyfarth et al., 2003; Hackert, 2003; Fischer and Blickhan, 2006). The instant of touch-down of the limbs is tuned precisely by the so-called pre-stance retraction of the trailing limb, i.e. the movement of rotation of the trailing limb toward the ground before contact with it. This movement of rotation is a general characteristic of the kinematics of bouncing and running: during the last instants of the swing phase, the shoulder blade initiates a back rotation of the forelimb in many quadruped mammals (Fischer et al., 2002). Because of the requirement for precision in the motion of the single limb, asymmetrical locomotion could be considered to be a high level task.

Moreover, spatiotemporal coordination between the limbs of the pair (and between the pairs) makes asymmetrical locomotion closer to bimanual manipulative tasks than unimanual task. The mean values of ABS-HI in non human primates during bimanual manipulation are higher (ABS-HI=0.80 in a male group of capuchin monkeys, *Cebus apella*, and 0.76 in a mixed group of white-faced capuchin monkeys, *Cebus capucinus*) than during unimanual task, for example a reaching movement [mean ABS-HI=0.4 in Spinozzi et al (1998) and 0.31 in Meunier and Vauclair (2007); note that the value for reaching movements in primates and dogs are comparable]. Involving the familiar and repetitive aspects of galloping but also a requirement for spatiotemporal precision and limb coordination, asymmetrical locomotion comes between low and high level tasks as defined by Fagot and Vauclair (1991). Moreover, although a dynamic stable gallop requires a threshold speed, an increase of speed only weakly enhances dynamic stability (Seyfarth et al, 2002). In Bernstein's view, a fast gallop is as complex as a slow gallop, at least in the context of advancing in a straight line on even terrain, because the number of degrees of freedom of the musculo-skeletal system to be controlled is the same (Bernstein, 1984). Consistent with this view, we found that the strengths of preference were independent of speed during galloping.

Asymmetrical gaits are a quasi exclusivity of mammals. The comparison of the morphology of the oldest mammalian fossils with the morphology of extant small mammals indicates that the potential for such asymmetrical locomotion was already present in early mammals (Ji et al., 2002). The introduction of asymmetrical locomotion with a shift between forelimbs (gallop, half-bound) minimized the vertical displacement of the centre of mass and thus led to more economic locomotion (Hackert, 2003). Moreover, the functional differentiation between trailing and leading limb during asymmetrical locomotion gives rises to a limb preference that may also reduce energetic cost. Indeed, torques that act in the preferred arm during manipulative movements in humans are less than those in the other arm, as both arms performed the same task by using the same trajectory (Bagesteiro and Sainburg, 2002). These observations lead us to suggest that limb preference was present in the locomotion of early mammalian species. The ability to perform precise and/or powerful manipulations may have sustained the lateralization of brain organization through evolution (Vallortigara et al., 1999; Rogers and Andrew, 2002; Vallortigara and Bisazza, 2002; Bradshaw and Rogers, 1993), so it is plausible that the benefits of an asymmetrical gait for efficient locomotion may also have sustained lateralization.

## Acknowledgments

Observations of pikas were funded by the DFG (Deutsche Forschungsgemeinschaft) at the college of innovation "Motion Systems" at the Institute of Systematic Zoology of the University of Iena. Observations of dogs were funded by the French ANR (Agence Nationale pour la Recherche) as part of the LOCOMO project (ANR 06 BLAN 0132-02). We are indebted to Dr. Alain Puget (IPBS), Toulouse, for providing us with pikas and to the 132e BCAT of the French Army for its collaboration in the study; in particular we thank Lt.-Cl. Deuwel, Cdt Lamour, Cpl.-Ch. Séné, Grardel, Debiasi and Lievens, 1st Cl. Noroy, Baron and Baud.

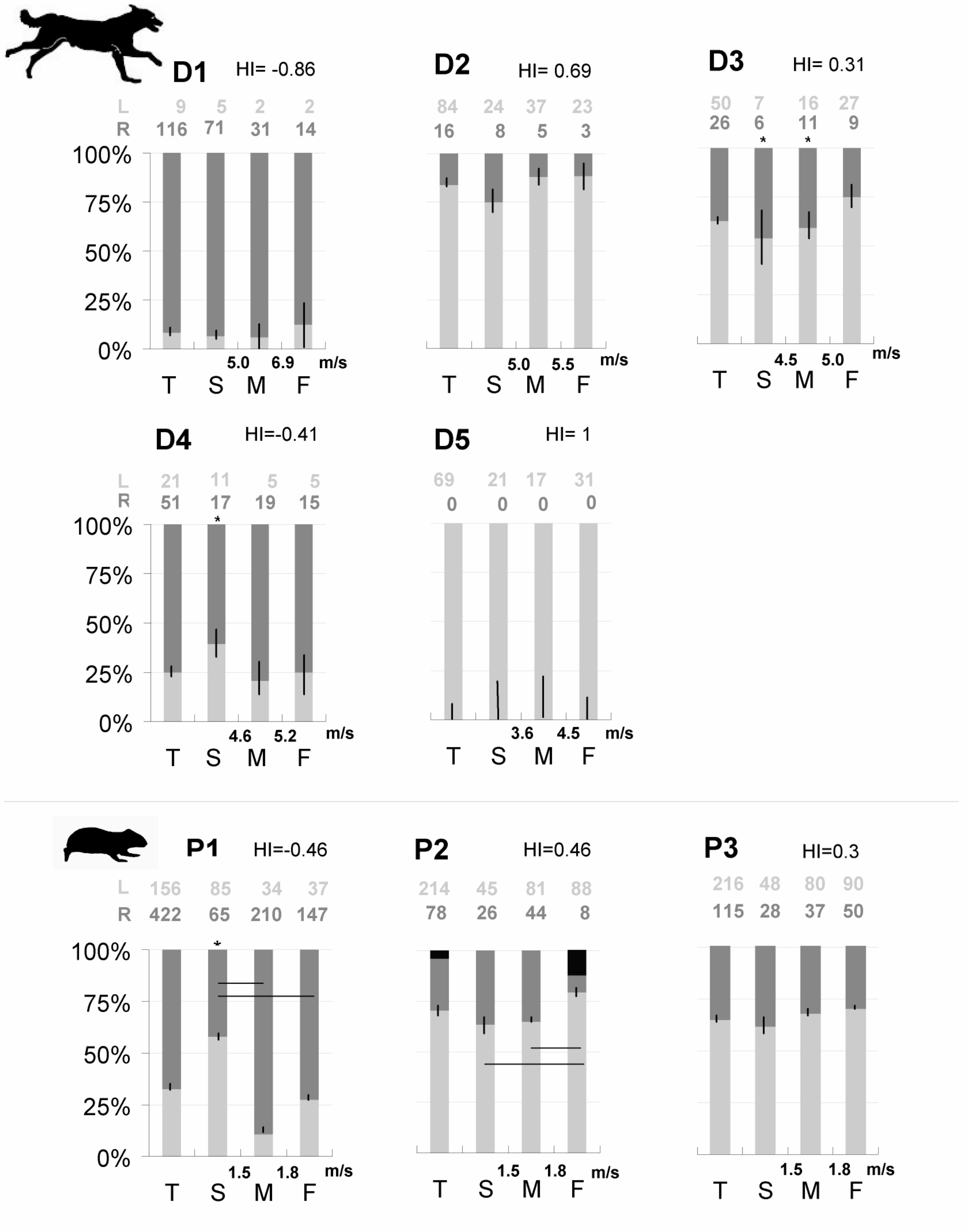

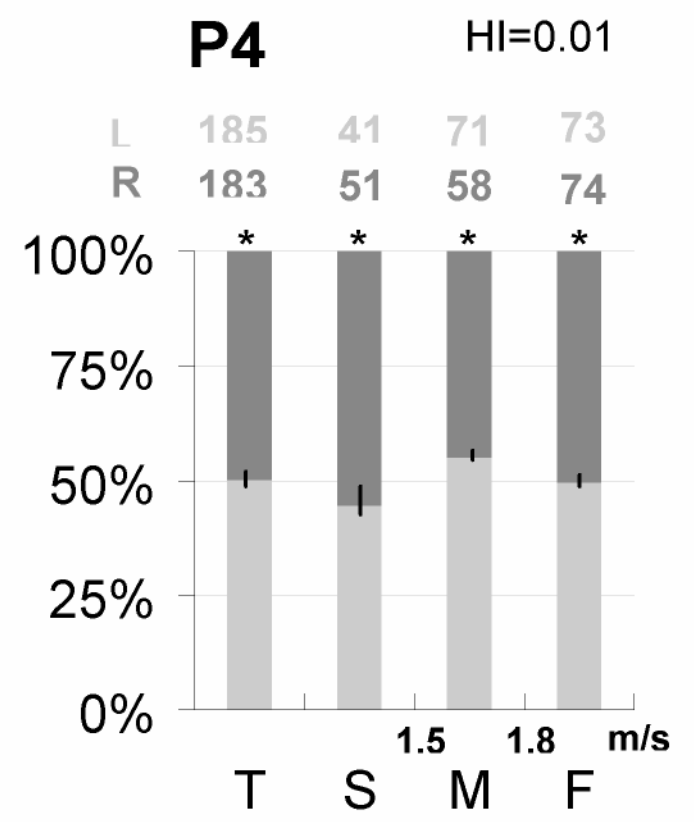

**Fig. 1**: **L**eft trailing (light grey) / **R**ight trailing (**dark grey**) distributions in 5 dogs (D) and 4 pikas (P). In a small number of sequences pika 2 synchronized both forelimbs (in black). Locomotor cycles are classified according to speed as: slow (S), medium (M) and fast (F) gallop or half-bound. The T class included all the cycles. The limits of speed classes are given below the horizontal axis. Black lines inside each class represent the truncation error (uncertainty). Stars (*) label distributions that are not significantly different from equidistribution and horizontal black lines connect pair of speed classes with significantly different distributions. For each speed class the left/right trailing distribution (raw data) is given above the histogram.